\documentstyle[12pt]{article}
\begin{document}
\thispagestyle{empty}

\newcommand{\be}{\begin{equation}}
\newcommand{\ee}{\end{equation}}
\newcommand{\sect}[1]{\setcounter{equation}{0}\section{#1}}
\renewcommand{\theequation}{\thesection.\arabic{equation}}
\newcommand{\vs}[1]{\rule[- #1 mm]{0mm}{#1 mm}}
\newcommand{\hs}[1]{\hspace{#1mm}}
\newcommand{\mb}[1]{\hs{5}\mbox{#1}\hs{5}}
\newcommand{\bea}{\begin{eqnarray}}
\newcommand{\ena}{\end{eqnarray}}

\newcommand{\wt}[1]{\widetilde{#1}}
\newcommand{\und}[1]{\underline{#1}}
\newcommand{\ov}[1]{\overline{#1}}
\newcommand{\sm}[2]{\frac{\mbox{\footnotesize #1}\vs{-2}}
		   {\vs{-2}\mbox{\footnotesize #2}}}
\newcommand{\prt}{\partial}
\newcommand{\eps}{\epsilon}

\newcommand{\R}{\mbox{\rule{0.2mm}{2.8mm}\hspace{-1.5mm} R}}
\newcommand{\Z}{Z\hspace{-2mm}Z}

\newcommand{\cd}{{\cal D}}
\newcommand{\cg}{{\cal G}}
\newcommand{\ck}{{\cal K}}
\newcommand{\cw}{{\cal W}}

\newcommand{\vj}{\vec{J}}
\newcommand{\vl}{\vec{\lambda}}
\newcommand{\vz}{\vec{\sigma}}
\newcommand{\vt}{\vec{\tau}}
\newcommand{\vw}{\vec{W}}
\newcommand{\poiss}{\stackrel{\otimes}{,}}

% REVUES POUR BIBLIO

\newcommand{\NP}[1]{Nucl.\ Phys.\ {\bf #1}}
\newcommand{\PL}[1]{Phys.\ Lett.\ {\bf #1}}
\newcommand{\NC}[1]{Nuovo Cimento {\bf #1}}
\newcommand{\CMP}[1]{Comm.\ Math.\ Phys.\ {\bf #1}}
\newcommand{\PR}[1]{Phys.\ Rev.\ {\bf #1}}
\newcommand{\PRL}[1]{Phys.\ Rev.\ Lett.\ {\bf #1}}
\newcommand{\MPL}[1]{Mod.\ Phys.\ Lett.\ {\bf #1}}
\newcommand{\BLMS}[1]{Bull.\ London Math.\ Soc.\ {\bf #1}}
\newcommand{\IJMP}[1]{Int.\ Jour.\ of\ Mod.\ Phys.\ {\bf #1}}
\newcommand{\JMP}[1]{Jour.\ of\ Math.\ Phys.\ {\bf #1}}
\newcommand{\LMP}[1]{Lett.\ in\ Math.\ Phys.\ {\bf #1}}

%\begin{document}
\renewcommand{\thefootnote}{\fnsymbol{footnote}}

\newpage
\setcounter{page}{0}
\pagestyle{empty}

\vs{30}

\begin{center}

{\LARGE {\bf On Classical Rational ${\cw}$ Algebras}}\\[1cm]

\vs{10}

\vs{10}

{\large {F. Toppan}}\\
\quad \\
{\em Laboratoire de Physique Th\'eorique }\\
{\small E}N{\large S}{\Large L}{\large A}P{\small P}\\
{\em ENS Lyon, 46 all\'ee d'Italie,} \\
{\em F-69364 Lyon Cedex 07, France.}

\end{center}
\vs{20}

\centerline{ {\em  Talk given at Rakhov (Ukr), February 1994}}

\vs{20}

\centerline{ {\bf Abstract}}

\indent

The structure of classical non-linear $\cw$ algebras closing on
rational
functions is analyzed both for the ordinary and the supersymmetric
case.
Such algebras appear as a result of a coset construction. Their
relevance
to physical applications is pointed out.

\newpage
\pagestyle{plain}
\renewcommand{\thefootnote}{\arabic{footnote}}
\setcounter{footnote}{0}

\sect{Introduction}

\indent

The purpose of this talk is to illustrate the basic properties of the
so-called classical rational $\cw$ algebras which have been
introduced in \cite{DFSTR}
and further developped in \cite{ftoppan}. \par
The name of rational $\cw $ algebra is deserved to a new class of
$\cw$ algebras, involving a  finite set of generators, among them a
stress-energy
tensor, while the other fields are primary or quasi-primary with
respect to the latter. The algebra is introduced through a Poisson
brackets structure among the generators: it satisfies the standard
properties of antisymmetry, linearity and Jacobi identities. While
standard Kac-Moody algebras close in a linear way, and
standard (polynomial) $\cw$ algebras, non-linearly but in terms of
polynomials
in the generators and their derivatives, rational $\cw$ algebras have
Poisson
brackets of the following kind
\bea
\{ W_i (z),W_j (w) \} &=& \sum_{k=1,..N} F_k (W_r (w),\partial^n\
 W_r (w) ) \delta^{(k)} (z-w)
\label{properties}
\ena
where $F_k$ are rational functions (quotient of homogeneous
polynomials) in the generators $W_r$ and their derivatives.\par
Algebras satisfying the above property appear when considering coset
constructions, i.e. the factorization of a spin $1$-fields Kac-Moody
subalgebra
out of a given algebra. Such statement will provide the spin $1$
extension of the factorization theorem concerning spin ${\textstyle
{1\over 2}}$ fields due to
Goddard and Schwimmer (see \cite{GS}). \par
Constructions different from ours, leading to algebras
satisfying the above (\ref{properties}) property have been discussed
in \cite{feher}.
Their quantum extension have been analyzed in \cite{honecker}. \par
In this talk I will review the basic properties of bosonic and
supersymmetric rational $\cw$ algebras. At the end I will discuss
their relevance in some applications, mainly concerning the hierarchy
of integrable equations.

\sect{Bosonic Rational $\cw$ Algebras}

\indent

In this section the construction of rational $\cw$
algebras will be reviewed. For simplicity only
the abelian (${\hat{U(1)}}$) quotients will be considered.

Let ${\hat{{\cal G}}}$ or $\cw$ denote respectively a Kac-Moody or a
$\cw$
algebra
admitting a subalgebra generated by a ${\hat {U(1)}}$ Kac-Moody
current
$J(z)$:
\be
\{J(z), J(w)\} = \gamma \delta'(z-w)\equiv \gamma\prt_w\delta(z-w)
\ee
(in the classical case the normalization factor $\gamma$ can be fixed
without loss of generality as will be done in the following).
It is possible to express any other element of the algebra
in terms of a basis of fields $W_{i,q_i}$ having a definite charge
$q_i$ with respect to $J(z)$, namely satisfying:
\be
\{J(z), W_{i,q_i}(w)\} = q_i W_{i,q_i}\delta(z-w)
\ee
A derivative $\cd$, covariant with respect to the above relations,
can be introduced (see also \cite{bai}):
\be
\cd W_{i,q_i}(w) = (\prt -{\textstyle{q_i\over \gamma}} J(w))
W_{i,q_i}(w)
\ee
The elements in $Com_J({\cal G},\cw)$ form the subalgebra of the
enveloping algebra commuting with $J(w)$; \footnote{ It is better
conceptually to understand the derivative operator $\prt
={\textstyle{d\over dw}}$ as an element of the original algebra, as
this is the case for Kac-Moody algebras.} they are spanned by
the vanishing total charge monomials \\
$(\cd^{n_1}V_{1,q_1})(\cd^{n_2}V_{2,q_2})...
(\cd^{n_j}V_{j,q_j})$, where the $n_i$'s are non negative integers
and the total
charge is $q=q_1+q_2+...q_j =0$.\\
{}From now on we will concentrate only on invariants produced by
bilinear combinations such as
\begin{eqnarray}
\cd^pV_+\cdot\cd^qV_-&&
\label{bilin}
\end{eqnarray}
(with $p,q\geq 0$ and $V_\pm$ have opposite charges),
together with of course originally invariant fields. This is still a
closed
algebra. Let me summarize the basic results of \cite{DFSTR}, with
some extra
comments:\\
{\it i}) there exists a linear basis of fields, given by $V^{(p)}=
\cd^p V_+\cdot V_- $ such that any bilinear invariants of the kind
(\ref{bilin}) is a linear combination
of the $V^{(p)}$'s and the derivatives acting on them.\\
{\it ii}) the Poisson brackets algebra of the fields $V^{(p)}$'s
among themselves is
closed (possibly with the addition of other invariants, in the
general case),
but never in a finite way (the Poisson brackets of $V^{(p)}$ with $
V^{(q)}$ necessarily generates on the right hand side terms depending
on $V^{(p')}$, with $p'> p,q $). Moreover it can be explicitly
checked
that it is a non-linear algebra, so that it has the structure of a
non-linear $W_\infty$
algebra.\\
{\it iii}) due to the properties of the covariant derivative, the
fields $V^{(p)}$, which are linearly independent, satisfy algebraic
relations like the following quadratic ones
\bea
V^{(p+1)} \cdot V^{(0)} &=& V^{(0)} \cdot \partial V^{(p)}
+V^{(p)}\cdot ( V^{(1)} -
\cdot
\partial V^{(0)})
 \label{rela}
\ena
Such relations allow to express algebraically the fields $V^{(p)}$,
for $p\geq 2$ in terms of the fundamental fields $V^{(0)}$ and
$V^{(1)}$. The above derived non-linear $\cw_\infty$ algebra has
therefore the structure of a rational $\cw $ algebra. Notice that
relations like (\ref{rela}) contain no informations if the fields
$V_\pm$ are fermionics. (\ref{rela}) can still be applied to
superalgebras if $V_\pm$ are bosonic superfields.
\\
{\it iv}) if $Com_J({\cal G}, \cw)$ contains a field $T(w)$ (the
stress-energy tensor)
whose Poisson brackets are the Virasoro algebra with non-vanishing
central charge, and moreover $V^{(0)}$ is primary with conformal
dimension $h$, then there exists a one-to-one correspondence between
the fields $V^{(p)}$ of the basis, and an infinite tower of uniquely
determined fields $W_{h+p}$, primary with respect to $T$, with
conformal dimension $h+p$. This relation should be understood as
follows: $V^{(p)}$ is the leading term in the associated primary
field. The remaining terms are fixed without ambiguity, some of
them just requiring $W_{h+p}$ being primary, some others once
a specific scheme to determine them is adopted (as an analogy, one
should think to the choice of the renormalization scheme when dealing
with renormalizable quantum field theories).\\
 As we will see, the condition of having a
non-vanishing central charge drops for the
${{\hat{sl(2)}}}/{\hat{U(1)}}$ coset model:
therefore no infinite tower of primary fields associated to
each $V^{(p)}$ can be generated
(we have an infinite tower of ``almost" primary fields
associated to them). An infinite number of primary fields can still
be
produced, but they are of a trivial type, being just products of
lower
order primary fields.
 Anyway the
structure of
a rational algebra with two primary fields is mantained even in
this case.  The next
simplest model admitting an infinite tower of invariant primary
fields associated to $V^{(p)}$ is based on the coset
${\textstyle {{\hat{sl(2)}}\times {\hat{U(1)}}\over {\hat{U(1)}}}}$,
since now there
exists another ${\hat{U(1)}}$ current,commuting with $J(w)$, which
allows to
define an invariant stress-energy tensor with non-vanishing charge.

To be definite let us discuss here the simplest example, given by
the coset
 ${\textstyle{ {\hat{sl(2)}}\over {\hat{U(1)}}}}$.

The classical ${\hat{sl(2)}}$ algebra is given by the following
Poisson brackets:
\bea
\{J_+(z),J_-(w)\} &=&  \delta'(z-w) - 2 J_0 (w) \delta(z-w) \equiv
\cd (w)\delta(z-w) \nonumber \\
 \{J_0(z), J_\pm (w)\} &=& \pm J_\pm (w) \delta (z-w)
 \nonumber \\
 \{J_0 (z),J_0(w)\} &=& {-\textstyle{1\over
 2}}\delta'(z-w)\nonumber\\
\{J_\pm(z),J_\pm(w)\} &=& 0
\label{KM}
\ena
 $J_\pm$ play the role here of the fields $V_\pm$;
 they have conformal dimension $1$
(here and in the following, the symbol $\delta'(z-w)$ is understood
as $\prt_w\delta (z-w)$).\\
The rational coset algebra of the commutant with respect to the $J_0$
current is given by the following Poisson brackets
\bea
\{W_2(z),W_2(w)\} &=& 2 W_2(w) \delta'(z-w) + \prt W_2(w) \delta(z-w)
\nonumber \\
\{W_2(z),W_{3}(w)\} &=& 3 W_{3}(w) \delta'(z-w) + \prt
W_{3}(w) \delta(z-w) \quad \nonumber \\
\{W_3(z),W_3(w)\} &=& 2 W_2(w) \delta'''(z-w) + 3 \prt W_2(w)
\delta(z-w)'' +\nonumber\\
&& [16 V^{(2)} - 8 \prt W_3 +8 {W_2}^2 -3 \prt^2 W_2 ] (w)
\delta'(z-w)+
\nonumber \\
&&\prt_w [8 V^{(2)} - 4 \prt W_3 +4 {W_2}^2 -2\prt^2 W_2 ] (w)
\delta(z-w)\nonumber\\
&&
\label{coset}
\ena
We have preferred to express the above algebra in the basis of
(uniquely determined) primary fields
$W_2 = J_+\cdot J_-$ and $W_3 = \cd J_+ \cdot J_- - J_+ \cdot \cd J_-
$.
They have dimension $2,3$ respectively, while
\bea
V^{(2)} &=& \cd^2 J_+ \cdot J_- = {1\over 4 W_2}[ W_3^2 +2 W_2\prt
W_3 + 2 W_2\prt^2W_2 -\prt W_2\prt W_2].
\ena
The second equality follows from the relation (\ref{rela}).\\
$W_2$ plays the role of a stress-energy tensor having no central
charge.
As already stated, in this simple example there exists no infinite
tower of primary fields associated to
the $V^{(p)}$'s fields, the only primary ones being $W_{2,3}$
and their products ${W_2}^m {W_3}^n$ for $m,n$ non-negative
integers.

The algebra of the fields $V^{(p)} = \cd^p J_+\cdot J_-$ is a
non-linear $\cw_\infty$ algebra: if we let from the very beginning
identify $J_0\equiv 0$,
then $J_\pm$ can be identified with the fields $\prt \beta$ and
$\gamma$ of a bosonic $\beta -\gamma$ system, the covariant
derivative in $V^{(p)}$ must be replaced by the ordinary derivative
and the non-linear $\cw_\infty$ algebra is reduced to the standard
linear $w_\infty$ algebra.

\section{The $N=1$ supersymmetric case}

\indent

In this section I will extend the definition of rational $\cw$
algebras to
the $N=1$ supersymmetric case, which presents as already pointed out
some peculiarities with respect to the bosonic case. Examples
concerning the
supersymmetric case can be found also in \cite{DFSR}. For simplicity
the discussion will be limited to the abeilan coset.\par
The $N=1$ superspace is introduced through the supercoordinate
$X\equiv, x, \theta$, with $x$
and $\theta
$ real, respectively bosonic and grassmann, variables. The
supersymmetric spinor
derivative is given by
\bea
D \equiv D_X &=& {\partial\over \partial\theta} +\theta
{\partial\over \partial x}
\ena
(therefore $ {D_X}^2 ={\textstyle{\partial\over
\partial x}}$). \par
The supersymmetric delta-function $\Delta (X,Y)$ is a fermionic
object
\bea
\Delta (X,Y) &=& \delta (x-y) (\theta -\eta)
\ena
In order to produce covariant derivatives we have therefore to look
for a fermionic spin ${\textstyle{1\over 2}}$ superfield which is the
$N=1$ counterpart of the $U(1)-{\cal KM}$ current $J_0(z)$. It can be
introduced as $\Psi_0(X)= \psi_0 (x) +
\theta J_0(x)$, satisfying the super-Poisson brackets
relation\footnote{
we recall that super-Poisson brackets are symmetric  when taken
between odd elements, antisymmetric otherwise.}
\bea
\{ \Psi_0 (X), \Psi_0 (Y) \} &=& D_Y \Delta (X,Y)
\label{zerosusyalg}
\ena
which implies, at the level of components
\bea
\{\psi_0(x),\psi_0(y)\}&=&-\delta (x-y)\nonumber\\
\{J_0(x),J_0(y)\}&=&-\partial_y\delta (x-y)
\ena
Super-covariant fields and the supercovariant derivative can now be
defined
through the relations
\bea
\{ \Psi_0 (X), \Phi_q (Y) \} &=& q\Delta (X,Y) \Phi_q (Y)\nonumber\\
{\cal D}\Phi_q &=& D\Phi _q + q \Psi_0 \Phi_q
\ena
$\Phi_q$ is a covariant superfield (either bosonic or fermionic).\par
Let us specialize now our discussion to the simplest example of
supersymmetric algebra giving rise to a rational coset: it is
introduced in terms of
 two spin ${\textstyle{1\over 2}}$
superfields $\Psi_\pm $ superfields which can be assumed to be either
fermionic
or bosonic (fermionic in the following discussion). The Poisson
brackets are given by the relations
\bea
\{ \Psi_0 (X), \Psi_\pm (Y) \} &=& \pm \Delta (X,Y) \Psi_\pm
(Y)\nonumber \\
\{ \Psi_+ (X), \Psi_- (Y) \} &=& {\cal D}_Y \Delta (X,Y) = D_Y \Delta
(X,Y)
+ \Delta (X,Y) \Psi_0(Y)
\nonumber\\
\{ \Psi_\pm (X), \Psi_\pm (Y) \} &=& 0
\label{susyalg}
\ena
The linear generators of the commutants are the composite superfields
$V_n (X)$,
\bea
V_n &=& \Psi_- {\cal D}^{n} \Psi_+ \quad\quad n=0,1,2,...
\label{superinv}
\ena
which have vanishing Poisson brackets with respect to
$\Psi_0 $:
\bea
\{ \Psi_0 (X), V_n (Y) \} &=& 0
\ena
The superfields $V_n$ are respectively bosonic for even values of $n$
and fermionic for odd values.
The set
$\{V_0, V_1\}$
constitutes a finite super-algebra, given by the Poisson brackets
\bea
\{ V_0 (X), V_0(Y) \} &=& -\Delta (X,Y) (DV_0 +2V_1)(Y)\nonumber\\
\{ V_0 (X), V_1(Y) \} &=& \Delta^{(2)} (X,Y))
V_0(Y) +\Delta^{(1)}(X,Y) V_1(Y) -\Delta (X,Y) DV_1 (Y)\nonumber\\
\{ V_1 (X), V_1(Y) \} &=& -2\Delta^{(2)} (X,Y) V_1(Y) -\Delta (X,Y)
{D_Y}^2V_1(Y)
\label{supcos}
\ena
In terms of component fields it is given by two bosons of spin $1$
and $2$
respectively, and two spin ${\textstyle {3\over 2}}$ fermions. It is
the maximal
finite subalgebra of the coset superalgebra: as soon as any other
superfield is added
to $V_0,V_1$, the whole set of fields $V_n$ is needed to close the
algebra, giving to the coset the structure of a super-$\cw_\infty$
algebra. Moreover such algebra closes in non-linear way.
Such a superalgebra is associated to the $N=1$ supersymmetric
non-linear Schr\"{o}dinger equation. \par
Let us make now some comments concerning the rational
character of
the above defined super-$\cw_\infty$ algebra: the whole set of
algebraic relations can be
expressed just in terms of closed rational super-$\cw$ algebra
involving $4$ superfields as the following reasoning shows: let us
introduce the superfields
\bea
\Lambda_p &=_{def}& {\cal D} \Psi_- \cdot {\cal D}^{(p+1)}
\Psi_+\nonumber
\ena
then
\bea
\Lambda_p &=& D V_{p+1} - V_{p+2} \nonumber
\label{lambdadef}
\ena
Due to standard properties of the covariant derivative we can write
down for the superfields $\Lambda_p$ the analogue of the relation
(\ref{rela}) of the bosonic case:
\bea
\Lambda_0 \Lambda_{p+1} &=& \Lambda_0 D\Lambda_p + (\Lambda_1 -
D\Lambda_0)\Lambda_p
\label{ratio2}
\ena
which implies that $\Lambda_p$ are rational functions of
$\Lambda_{0,1}$, which
in their turns are determined by $V_i$, $i=0,1,2,3$. \\
Inverting the relation (\ref{lambdadef}) we can express any higher
field $V_{p+1}$ in terms of $V_p$, $\Lambda_{p-1}$. As a consequence
of this we have
the (rational) closure of the superalgebra on the superfields
$V_0,V_1, V_2, V_3$. \par
We remark that now is not possible, like in the bosonic case, to
determine higher
order superfields $V_p$ from the formula (\ref{ratio2}) by simply
inserting $V_0$, $V_p$ in place of $\Lambda_0$, $\Lambda_p$: this is
due to the fact that
any product $V_0\cdot V_{p+1} $ identically vanishes since it is
proportional to a squared  fermion (${\Psi_-}^2=0$).  That is the
reason why four superfields are
necessary to produce a finite rational algebra and not just two as
one would
have nively expected.
{}~\quad\\
\vfill
\newpage
{\Large {\bf Conclusions}}

\indent

In this talk I have analyzed the mathematical structure of rational
$\cw$ algebras, both for the bosonic and the supersymmetric case,
and I have provided a general framework to construct such kind of
algebras.\par
In this conclusion I feel necessary to provide some insights
about why
such structures are relevant for physical applications.
The first problems which could be mentioned are those, like the
quantum Hall
effect or the black hole problem \cite{bakas} which involve a
$\cw_\infty$ algebra. Even if rational algebras have not been
explicitly exploited in such
problems, it is very likely they play a role. This is certainly sure
for
the black hole problem, where the appearance of a rational $\cw$
algebra is a well-established fact. There is however another field of
applications of
rational $\cw$ algebras which is particularly interesting and well
understood now \cite{ftoppan}. It concerns the hierarchies of
integrable
equations and their relations to matrix models. \\
Such hierarchies can be regarded as consistent reductions of KP and
super-KP
flows. In the physical literature \cite{aratyn}
appeared recently non-standard reductions
of the KP-hierarchy, called multi-field reductions, which have
different
properties with respect to the standard Drinfeld-Sokolov type of
reductions.
In particular they do not lead to a purely differential Lax operator.
It turns
out that, while standard Drinfeld-Sokolov reductions are related to
polynomial $\cw$ algebras, multifield reductions are related to
rational (coset) $\cw$ algebras. Better stated, there exists a
Poisson brackets structure for the
reduced KP hierarchy, such that the hamiltonian densities of the
infinite
tower of hamiltonians in involution belong all to the coset algebra
and
therefore generate a rational $\cw$ algebra. \par
The simplest example of such kind hierarchy is given by the
Non-Linear-Schr\"{o}dinger equation, which is associated to the
previously
analyzed ${\textstyle{\hat{sl(2)}\over{\hat{U(1)}}}}$ coset. The
supersymmetric
rational $\cw$ algebra discussed above is related to the $N=1$
extension of the
above equation. More complicated cosets provide new integrable
hierarchies
and KP reductions.

{}~\\~\\

\noindent
{\large{\bf Acknowledgements}}
{}~\\~\\
It is a pleasure to thank for useful discussions
F. Delduc,
L. Frappat, E. Ragoucy and P. Sorba.
{}~\\
{}~\\

\end{document}